\documentclass[amsmath,amssymb,aps,10pt,prd,letterpaper,nofootinbib,balancelastpage,notitlepage,twocolumn,floatfix]{revtex4-2}
\usepackage{graphicx}	% Standard figures 
\usepackage{ragged2e}	
\usepackage{bm}		    % Bold math
\usepackage[sort&compress]{natbib}	 	% Standard reference package
	\setcitestyle{square,numbers,comma}	% Set citation style
\usepackage[colorlinks=true,urlcolor=blue,linkcolor=black,citecolor=red]{hyperref}	
\usepackage{xcolor}
\usepackage{fontawesome5}
\definecolor{orcidlogocol}{rgb}{0.65, 0.807, 0.223}
\newcommand{\orcid}[1]{\,\href{https://orcid.org/#1}{\textcolor{orcidlogocol}{\footnotesize\faOrcid}}}
\newcommand{\tsc}[1]{\textsc{#1}}
\newcommand{\mc}[1]{\ensuremath\mathcal{#1}}
\newcommand{\LL}{\mathcal{L}}
\newcommand{\dm}{\textsc{dm}}
\newcommand{\sinc}{\ensuremath{\mathrm{sinc}}}
\newcommand{\tabref}[2][]{Tab{#1}.~\ref{#2}}		% Table reference
\newcommand{\figref}[2][]{Fig{#1}.~\ref{#2}}		% Figure reference
\newcommand{\sectref}[2][]{Sec{#1}.~\ref{#2}}		% Section reference
\newcommand{\appref}[2][x]{Appendi{#1}~\ref{#2}}		% Section reference
\renewcommand{\eqref}[2][]{Eq{#1}.~(\ref{#2})}		% Equation reference
\newcommand{\eqrefRange}[2]{Eqs.~(\ref{#1})--(\ref{#2})}		% Equation range reference
\newcommand{\citeR}[2][]{Ref{#1}.~\cite{#2}}			% Ref. Citation
\newcommand{\lb}{\ensuremath{\left}}					% Left Brackets
\newcommand{\rb}{\ensuremath{\right}}					% Right Brackets
\newcommand{\order}[1]{\ensuremath{\mathcal{O}(#1)}}    % O(x) notation
\begin{document}
%%%%%%%%%%%%%%%%%%%%%%%%%%%%%%%%%%%%%%%%%%%%%%%%%%%%%%%%%%%%%%%%%%%%%%%%%%%%%%%%%%%%%%%%%%	
%%%%%%%%%%%%%%%%%%%%%%%%%%%%%%%%%%%%%%%%%%%%%%%%%%%%%%%%%%%%%%%%%%%%%%%%%%%%%%%%%%%%%%%%%%	
%%%%%%%%%%%%%%%%%%%%%%%%%%%%%%%%%%%%%%%%%%%%%%%%%%%%%%%%%%%%%%%%%%%%%%%%%%%%%%%%%%%%%%%%%%	
%%%%%%%%%%%%%%%%%%%%%%%%%%%%%%%%%%%%%%%%%%%%%%%%%%%%%%%%%%%%%%%%%%%%%%%%%%%%%%%%%%%%%%%%%%

%%%%%%%%%%%%%%%%%%%%%%%%%%%%%%%%%%%%%%%%%%%%%%%%%%%%%%%%%%%%%%%%%%%%%%%%%%%%%%%%%%%%%%%%%%	
% Title, Author and Affiliation
\title{Asteroids for ultralight dark-photon dark-matter detection}
\date{\today}
%%%%%%%%%%%%%%%%%%%%%%%%%%%%%%%%%%%%%%%%%%%%%%%%%%%%%%%%%%%%%%%%%%%%%%%%%%%%%%%%%%%%%%%%%%	

%%%%%%%%%%%%%%%%%%%%%%%%%%%%%%%%%%%%%%%%%%%%%%%%%%%%%%%%%%%%
\author{Michael A.~Fedderke\orcid{0000-0002-1319-1622}\,}
\email{mfedderke@jhu.edu}
\affiliation{The William H.~Miller III Department of Physics and Astronomy, The Johns Hopkins University, Baltimore, MD  21218, USA}
%%%%%%%%%%%%%%%%%%%%%%%%%%%%%%%%%%%%%%%%%%%%%%%%%%%%%%%%%%%%
\author{Anubhav Mathur\orcid{0000-0003-0973-1793}\,}
\email{a.mathur@jhu.edu}
\affiliation{The William H.~Miller III Department of Physics and Astronomy, The Johns Hopkins University, Baltimore, MD  21218, USA}
%%%%%%%%%%%%%%%%%%%%%%%%%%%%%%%%%%%%%%%%%%%%%%%%%%%%%%%%%%%%

%%%%%%%%%%%%%%%%%%%%%%%%%%%%%%%%%%%%%%%%%%%%%%%%%%%%%%%%%%%%%%%%%%%%%%%%%%%%%%%%%%%%%%%%%%
% Abstract
\begin{abstract}
%%%%%%%%%%%%%%%%%%%%%%%%%%%%%%%%%%%%%%%%%%
Gravitational-wave (GW) detectors that monitor fluctuations in the separation between inertial test masses (TMs) are sensitive to new forces acting on those TMs.
Ultralight dark-photon dark matter (DPDM) coupled to $U(1)_B$ or $U(1)_{B-L}$ charges supplies one such force that oscillates with a frequency set by the DPDM mass.
GW detectors operating in different frequency bands are thus sensitive to different DPDM mass ranges.
A recent GW detection proposal based on monitoring the separation of certain asteroids in the inner Solar System would have sensitivity to $\mu$Hz frequencies~[Fedderke \emph{et al}., \href{https://doi.org/10.1103/PhysRevD.105.103018}{Phys.~Rev.~D \textbf{105}, 103018 (2022)}].
In this paper, we show how that proposal would also enable access to new parameter space for DPDM coupled to $B$ [respectively, $B-L$] charges in the mass range $5 \ [9]\times 10^{-21}\,\text{eV} \lesssim m_{\dm} \lesssim 2 \times 10^{-19}\,\text{eV}$, with peak sensitivities about a factor of 500~[50] beyond current best limits on~$\varepsilon_B$\,[$\varepsilon_{B-L}$] at $m_{\dm} \sim 2\times 10^{-19}\,$eV.
Sensitivity could be extended up to $m_{\dm} \sim 2 \times 10^{-18}\,\text{eV}$ only if noise issues associated with asteroid rotational motion could be overcome.
%%%%%%%%%%%%%%%%%%%%%%%%%%%%%%%%%%%%%%%%%%
\end{abstract}
%%%%%%%%%%%%%%%%%%%%%%%%%%%%%%%%%%%%%%%%%%%%%%%%%%%%%%%%%%%%%%%%%%%%%%%%%%%%%%%%%%%%%%%%%%

\maketitle

%%%%%%%%%%%%%%%%%%%%%%%%%%%%%%%%%%%%%%%%%%%%%%%%%%%%%%%%%%%%%%%%%%%%%%%%%%%%%%%%%%%%%%%%%%
\section{Introduction}
\label{sect:intro}
%%%%%%%%%%%%%%%%%%%%%%%%%%%%%%%%%%%%%%%%%%%%%%%%%%%%%%%%%%%%%%%%%%%%%%%%%%%%%%%%%%%%%%%%%%
Dark matter (DM) constitutes 26\% of the energy budget of the Universe~\cite{Planck:2018vyg}.
Despite decades of increasingly sensitive search efforts, its fundamental identity remains a mystery.
Ultralight bosons are one interesting class of DM candidates.
They admit a classical-field description that oscillates in time at a frequency set by the DM mass.
In typical models, they are also very weakly coupled to the Standard Model (SM).
This permits sensitive experimental approaches targeting narrowband time-varying phenomena to shed light on the nature of the DM.

A compelling new-physics scenario involves gauging either the  $U(1)_B$ or $U(1)_{B-L}$ global symmetry of the SM, with the new gauge boson---the `dark' or `hidden' photon---coupling weakly to the associated current.%
%%%%%%%%%%%%
\footnote{\label{ftnt:otherDMcouplings}%
    A dark photon can also couple to the SM photon directly via the `vector portal', in which case it is referred to as the kinetically mixed dark photon; see, e.g., \citeR[s]{Holdom:1985ag,Nelson:2011sf,Arias:2012az}.} %
%%%%%%%%%%%%
The static, Weak Equivalence Principle (EP) violating, fifth-force effects induced by such couplings have been stringently constrained~\cite{%
% EotWash
Smith:1999cr,Schlamminger:2007ht,ADELBERGER2009102,Wagner:2012ui,Shaw:2021gnp,
% MICROSCOPE
Touboul:2017grn,Fayet:2017pdp,PhysRevLett.120.141101,PhysRevLett.129.121102}.%
%%%%%%%%%%%%%
\footnote{\label{ftnt:Yukawa}%
    Yukawa modifications to the $1/r^2$ gravitational force law also supply constraints, but these are typically weaker than the best direct EP-violation constraints in the DM mass range of interest in this work; see, e.g., \citeR{ADELBERGER2009102} and references therein.
    } %
%%%%%%%%%%%%
Sufficiently feebly coupled dark photons can however additionally serve as an ultralight bosonic DM candidate.
In this case, the oscillations of the dark-photon dark-matter (DPDM) field exert a minute oscillatory force on SM matter. 
In particular, such a force will cause the test masses (TMs) utilized in certain classes of gravitational-wave (GW) detectors to oscillate in a detectable fashion, allowing them to do double duty as DM detectors;%
%%%%%%%%%%%%
\footnote{\label{ftnt:massiveDMtoo}%
    Other DM candidates can also be searched for, in different ways: e.g., via the transient acceleration signals induced by supermassive DM states moving past the detector~\cite{Seto:2004zu,Adams:2004pk,Hall:2016usm,Baum:2022duc}.
    } % 
%%%%%%%%%%
the theory of this effect has been thoroughly developed~\cite{Arvanitaki:2014faa,Graham:2015ifn,Arvanitaki:2016fyj,Pierce:2018xmy,Morisaki:2018htj,Guo:2019ker,Morisaki:2020gui,LIGOScientificCollaborationVirgoCollaboration:2021eyz}.%
%%%%%%%%%%%%
\footnote{Note in particular Sec.~V.A.4 in the published version of \citeR{Graham:2015ifn}.} %
%%%%%%%%%%%%
See also, e.g., \citeR[s]{Stadnik:2014tta,Stadnik:2015xbn,Grote:2019uvn} for further discussion of the related case of scalar DM.

In this paper, we evaluate the DPDM detection prospects for the future GW detector concept based on direct ranging between certain inner Solar System asteroids that was recently proposed in \citeR{Fedderke:2021kuy} and that would have $\mu$Hz frequency sensitivity (see also \citeR{Fedderke:2020yfy} for a related noise study, and \citeR[s]{Sesana:2019vho,Bustamante-Rosell:2021daj,Blas:2021mpc,Blas:2021mqw,Fedderke:2022kxq} for other recent work on $\mu$Hz GW detection).

The rest of this paper is structured as follows: in \sectref{sect:signal}, we discuss the DPDM signal and relate it to GW detector sensitivity.
We present results, discuss them, and conclude in \sectref{sect:results}.
In \appref{app:kdetFactors}, we discuss some correction factors that are applied to published GW detector sensitivity curves in order to use them in this work.
In \appref{app:AveragingComment} we discuss how appropriate it is to consider results averaged over DPDM polarization and momentum orientations.

%%%%%%%%%%%%%%%%%%%%%%%%%%%%%%%%%%%%%%%%%%%%%%%%%%%%%%%%%%%%%%%%%%%%%%%%%%%%%%%%%%%%%%%%%%
\section{Signal and sensitivity}
\label{sect:signal}
%%%%%%%%%%%%%%%%%%%%%%%%%%%%%%%%%%%%%%%%%%%%%%%%%%%%%%%%%%%%%%%%%%%%%%%%%%%%%%%%%%%%%%%%%%
Consider gauging the SM $U(1)_{S}$ symmetry, where \linebreak$S\in\{B,B-L\}$, with that symmetry broken so as to give rise to a~(St\"uckelberg) mass $m_V$ for the associated $U(1)_S$ gauge boson $V_\mu$:
%%%%%%
\begin{align}
    \LL = \LL_{\tsc{sm}} - \frac{1}{4} V_{\mu\nu}V^{\mu\nu} + \frac{1}{2} m_{V}^2 V_\mu V^{\mu} - \varepsilon_S e_{\tsc{em}} V_\mu J_{S}^\mu\:,
    \label{eq:Lag}
\end{align}
%%%%%%
where $V_{\mu\nu} \equiv \partial_\mu V_\nu - \partial_\nu V_\mu$; $J_{S}^{\mu}$ is the relevant SM current; $e_{\tsc{em}} \equiv \sqrt{4\pi \alpha_{\tsc{em}}}$ is the fundamental EM charge unit, with $\alpha_{\tsc{em}}$ the electromagnetic~(EM) fine-structure constant; and~$\varepsilon_S$~parametrizes the $U(1)_S$ coupling strength to the DM, normalized to that of EM for $\varepsilon_S = 1$.
For $m_V \neq 0$, we necessarily have $\partial_\mu V^{\mu} = 0$.

The coupling $\epsilon_S\neq 0$ gives rise to a force on a SM object moving at speed $\bm{v}$:
%%%%%%
\begin{align}
    \bm{F} &= \varepsilon_S e_{\tsc{em}} Q_S \lb[ - \bm{\nabla} V^0 - \partial_t \bm{V} + \bm{v} \times ( \bm{\nabla}\times \bm{V} ) \rb]\:,
    \label{eq:force}
\end{align}
%%%%%%
where $Q_S$ is the $U(1)_S$ charge of the SM object.

If the $V_{\mu}$ gauge boson comprises all of the local, non-relativistic DM (speed $v_{\dm}\sim 10^{-3}$), we write $V^\mu \equiv (\mathcal{V}^0,\bm{\mathcal{V}})\exp\lb[ -i\omega_{\dm}t+i\bm{k}_{\dm}\cdot\bm{x}+i\alpha \rb]$ within a single coherence time/length,%
%%%%%%%%%%%%
\footnote{\label{ftnt:coherence}%
    The coherence time is $T_{\text{coh}} \sim 2\pi /(m_{\dm}v_{\dm}^2)$ and the coherence length is $\lambda_{\text{coh}} \sim 2\pi /(m_{\dm}v_{\dm})$ [i.e., the de Broglie wavelength].
    } 
%%%%%%%%%%%%
with $\alpha$ an arbitrary phase, $\omega_{\dm} \approx m_{\dm} \sqrt{ 1 + v_{\dm}^2 } \approx m_{\dm}$, and $|\bm{k}_{\dm}| \approx m_{\dm} v_{\dm}$.
Note that we have identified $m_{V} \equiv m_{\dm}$.
Now, $\partial_\mu V^{\mu} = 0$ implies that $|\mathcal{V}^0| \sim v_{\dm} |\bm{\mathcal{V}}|$.
Moreover, we have%
%%%%%%%%%%%%
\footnote{\label{ftnt:OOMests}%
    We clarify that these are order of magnitude estimates for the \emph{largest} that the expressions on the lhs can be; additional geometrical factors can suppress these even further.} %
%%%%%%%%%%%%
$|\bm{\nabla} \times \bm{V}| \sim m_{\dm} v_{\dm} |\bm{\mc{V}}|$ and $|\bm{\nabla} V^0| \sim m_{\dm} v_{\dm} |\mc{V}^0| \sim m_{\dm} v_{\dm}^2 |\bm{\mc{V}}|$.
Additionally, $|\partial_t \bm{V}| \sim m_{\dm} |\bm{\mc{V}}|$.
Since, by assumption, $T^{00}_V = \rho_{\dm}$, it follows that%
%%%%%%%%%%%%
\footnote{\label{ftnt:averaged}%
    Technically this is only true as an average statement over many coherence times of the DM field, as the field amplitude executes $\order{1}$ stochastic fluctuations from one coherence time to the next (see, e.g., \citeR[s]{Foster:2017hbq,Centers:2019dyn,Fedderke:2021aqo}); nevertheless, it is a good figure of merit.} %
%%%%%%%%%%%%
$|\bm{\mathcal{V}}| \sim \sqrt{2 \rho_{\dm}}/m_{\dm}$.

Considering objects gravitationally bound to the Solar System, we also have $|\bm{v}| \lesssim v_{\textsc{dm}}$, owing to the motion of the Solar System relative to the galactic rest frame.
Therefore, the force can be written as 
%%%%%%
\begin{align}
\begin{split}
    \bm{F} &\approx i \varepsilon_S e_{\tsc{em}} Q_S m_{\dm} \bm{\mc{V}} e^{-im_{\dm}(t-\bm{v}_{\dm}\cdot\bm{x})+i\alpha} \\ & \quad + \order{v_{\dm}^2,v_{\dm}|\bm{v}|}\times \varepsilon_S e_{\tsc{em}} Q_S m_{\dm} |\bm{\mc{V}}|\:.
    \label{eq:forceapprox}
\end{split}
\end{align}
%%%%%%

If the relevant SM object has mass $M$, and we drop the subleading corrections,%
%%%%%%%%%%%%
\footnote{\label{ftnt:subleading}%
    The vectorial orientation of the force relative to a GW detector baseline is the relevant quantity to consider for GW detector effects.
    Because the subleading corrections can be oriented differently than the leading term, it is possible that the leading term gives no effect, but that the subleading terms do.
    However, because the subleading corrections are suppressed by at least $\sim v_{\dm}^2$, this can only occur in highly tuned orientations.
    Real GW detector baselines evolve in orientation over time relative to inertial space, which will always spoil the tuning required for such cancellation to be maintained; see also \appref{app:AveragingComment}.
    The subleading terms can thus always be dropped.} %
%%%%%%%%%%%%
this force causes an oscillatory acceleration of that object~\cite{Graham:2015ifn,Pierce:2018xmy,Morisaki:2020gui,LIGOScientificCollaborationVirgoCollaboration:2021eyz}:
%%%%%%
\begin{align}
    \bm{a} &\approx 2 \varepsilon_S \sqrt{2\pi\alpha_{\tsc{em}} \rho_{\dm}}\, \frac{Q_S}{M} e^{-i m_{\dm} ( t - \bm{v}_{\dm} \cdot \bm{x} ) + i\phi}\, \bm{\hat{V}}\:,
    \label{eq:accn}
\end{align}
%%%%%%
where $\bm{\hat{V}}$ gives the polarization state of the DM ($\bm{\hat{V}} \cdot \bm{\hat{V}}^* = 1$), and we have absorbed a phase into $\phi$.

Consider now a GW detector that operates by measuring fluctuations in the proper distance between two or more inertial test masses that define the endpoints of one or more detector baselines.
The DM-induced acceleration \eqref{eq:accn} causes the TMs to oscillate, leading to a modulation of the proper length of the detector baseline(s) that manifests as an oscillatory strain component.
There are two contributions to this strain: (1) a term arising from the finite light-travel time for a null signal propagating between the TMs, present even if the acceleration had no spatial dependence at all (i.e., $v_{\dm} = 0$)~\cite{Arvanitaki:2014faa,Graham:2015ifn,Arvanitaki:2016fyj,Morisaki:2018htj,Morisaki:2020gui,LIGOScientificCollaborationVirgoCollaboration:2021eyz}, and (2) a term arising from the spatial gradient of the acceleration~\cite{Graham:2015ifn,Pierce:2018xmy,Guo:2019ker,Morisaki:2018htj,Morisaki:2020gui,LIGOScientificCollaborationVirgoCollaboration:2021eyz}.

Averaged over time, DM field orientations with respect to the baseline, and DM momentum directions, and ignoring further corrections at $\order{v_{\dm}^2}$, the mean square of the strain signal $h(t) \equiv \Delta t / (2L)$, where $\Delta t$ is the change to the round-trip light travel time between the TMs, can be written as~\cite{Morisaki:2020gui}:%
%%%%%%%%%%%%
\footnote{\label{ftnt:FPcavity}%
    For detectors that are constructed with Fabry-P{\'e}rot (FP) cavities in the baseline arms, these expressions are referred to the input in the sense that the FP transfer function has been omitted.
    Published GW characteristic-strain curves for detectors with FP cavities are similarly referred to the GW input.
    Since the FP cavity transfer function is the same for a GW and the DPDM case that we consider in this work, no correction is required for this~\cite{Morisaki:2020gui}; see also \citeR{Maggiore:2007zz} for further discussion of FP cavities in GW detectors.
	}%
%%%%%%%%%%%%
%%%%%%
\begin{align}
    \langle h^2 \rangle &= \langle h_1^2 \rangle + \langle h_2^2 \rangle \label{eq:hrms}\:;\\
    \langle h_1^2 \rangle &\equiv 4  c_1^{\text{geom}} \mathcal{H}^2 \times \sin^4\lb( \tfrac{1}{2} m_{\dm} L \rb) \label{eq:h1rms}\:;\\
    \langle h_2^2 \rangle &\equiv \frac{1}{3} c_2^{\text{geom}} \mathcal{H}^2 \times (m_{\dm}L v_{\dm})^2 \label{eq:h2rms}\:;\\
    \mathcal{H}^2 &\equiv \frac{8\pi}{3} \varepsilon_S^2 \alpha_{\tsc{em}} \frac{\rho_{\dm}}{m_{\dm}^4L^2} \lb( \frac{Q_S}{M} \rb)^2\:,\label{eq:HcalDefn}
\end{align}
%%%%%%
where $c_j^{\text{geom}}$ are $\order{1}$ geometrical factors that depend on the baseline orientation, and $L$ is the unperturbed baseline length.
The baseline is assumed to be approximately fixed at least on the $\sim 2L$ round-trip light travel time between the TMs, but it can vary secularly both in length and orientation on longer timescales.
The expression for~$\langle h_2^2 \rangle$ at \eqref{eq:h2rms} is correct in the limit $m_{\dm} L v_{\dm} \ll 1$; this is satisfied in all cases that we consider.
We discuss the appropriateness of averaging over DPDM field orientations and momentum directions at length in \appref{app:AveragingComment}, as well as a relevant correction we apply when we find this to be incompletely justified.

Note that $\langle h_2^2 \rangle / \langle h_1^2 \rangle \propto (m_{\dm} L v_{\dm})^2 \, \text{cosec}^{4}\lb( \frac{1}{2} m_{\dm} L \rb)$.
Because $m_{\dm}L v_{\dm}\ll 1$, the $h_2$ signal is only dominant when either (1) the baseline is much shorter than the DM Compton wavelength $m_{\dm} L \lesssim v_{\dm}$ (never the case for interesting mass ranges for the detectors we consider), or (2) in certain narrow mass ranges where the round-trip light travel time is such that the $h_1$ signal is nearly zero: $ |m_{\dm} L - 2\pi n| \sim 2\sqrt{m_{\dm}Lv_{\dm}}$ for $n=1,2,\ldots$.
The latter occurs only at frequencies above the peak sensitivity for every detector we consider.

The geometrical factors for a single-baseline detector are given by
%%%%%%
\begin{align}
c_1^{\text{geom}} = c_2^{\text{geom}} &= \frac{1}{2}\:,&& \text{[single baseline]} \label{eq:cGeomA}
\end{align}
%%%%%%
while for a two-baseline detector with an opening angle $\psi$ between the baselines, they are given by~\cite{Morisaki:2020gui}
%%%%%%
\begin{align}
c_1^{\text{geom}} &= 1-\cos\psi\:, &
c_2^{\text{geom}} &= 1-\cos^2\psi\:. \label{eq:cGeomB}
\end{align}
%%%%%%
For LIGO, $\cos\psi = 0$ (perpendicular arms), while for LISA, $\cos\psi \sim 1/2$ (equilateral triangle)~\cite{Morisaki:2020gui}.

It remains to relate the signal, \eqrefRange{eq:hrms}{eq:HcalDefn}, to a search sensitivity.
We work in the quasi-coherent, matched-filter approach of \citeR{Morisaki:2020gui} and, at a signal-to-noise ratio (SNR) of 1, estimate sensitivity using%
%%%%%%%%%%%%
\footnote{\label{ftnt:factors}%
    Some numerical and correction factors here differ from Eq.~(20) of \citeR{Morisaki:2020gui}.
    }%
%%%%%%%%%%%%
%%%%%%
\begin{align}
    \langle h^2\rangle \sim \frac{k_{\text{det}}(f_{\dm}) \times S_n\lb(f_{\dm}\rb)}{2 \times \min\lb[ T , \sqrt{T T_{\text{coh}}} \rb]}\:, \label{eq:sensitivity}
\end{align}
%%%%%%
where $T$ is the mission duration, $f_{\dm} \equiv m_{\dm}/(2\pi)$ is the frequency of the strain signal, $T_{\text{coh}} \sim 1/(f_{\dm} v_{\dm}^2)$ is the signal coherence time, and $\sqrt{ f S_n(f) } \equiv h_c(f)$ relates the one-sided strain-noise power spectral density~$S_n(f)$ to published characteristic-strain sensitivity curves~$h_c(f)$.
The factor $k_{\text{det}}(f)$ in \eqref{eq:sensitivity} is a further detector-specific, frequency-dependent correction factor~(see comments in \citeR{Morisaki:2020gui}) that accounts for (1) factors necessary to avoid double-counting of geometrical averaging effects, and (2) the fact that $h_c(f)$ [or $S_n(f)$] curves are conventionally specified with respect to a GW waveform input, while \eqrefRange{eq:hrms}{eq:HcalDefn} are specified with respect to the detector strain response.
We give detailed explanations and expressions for these factors in \appref{app:kdetFactors}.
Substituting \eqrefRange{eq:hrms}{eq:HcalDefn} and either \eqref{eq:cGeomA} or \eqref{eq:cGeomB} into \eqref{eq:sensitivity} gives the~$\varepsilon_S$ sensitivity for any detector of interest; we show these results in \sectref{sect:results}.

Changes in baseline length and/or orientation induce sidebands in the detector strain response that are relevant for the details of how a signal search would be conducted.
However, for our purposes, these effects can be ignored: the sensitivity curves that we employ do not vary much over a range $\Delta f \sim \Omega/(2\pi)$, where $\Omega$ is the orbital or rotational angular frequency of the baseline modulation.
Moreover, orbital- or rotational-velocity corrections to the time-averaged quantities $\langle h_{\{1,2\}}^2 \rangle$ are smaller than those displayed at \eqref[s]{eq:h1rms}~and~(\ref{eq:h2rms}): they would be at worst $\order{v_{\text{orbital}}^2}$, and $v_{\text{orbital}} \sim 0.1 v_{\dm}$ for TMs located at distances $\sim \text{AU}$ from the Sun (rotational effects for, e.g., detectors on Earth are even smaller).

%%%%%%%%%%%%%%%%%%%%%%%%%%%%%%%%%%%%%%%%%%%%%%%%%%%%%%%%%%%%%%%%%%%%%%%%%%%%%%%%%%%%%%%%%%
\section{Results and Discussion}
\label{sect:results}
%%%%%%%%%%%%%%%%%%%%%%%%%%%%%%%%%%%%%%%%%%%%%%%%%%%%%%%%%%%%%%%%%%%%%%%%%%%%%%%%%%%%%%%%%%

%%%%%%%%%%%%%%%%%%%%%%%%%%%%%%%%%%%
\begin{figure*}[t]
    \includegraphics[width=0.495\textwidth]{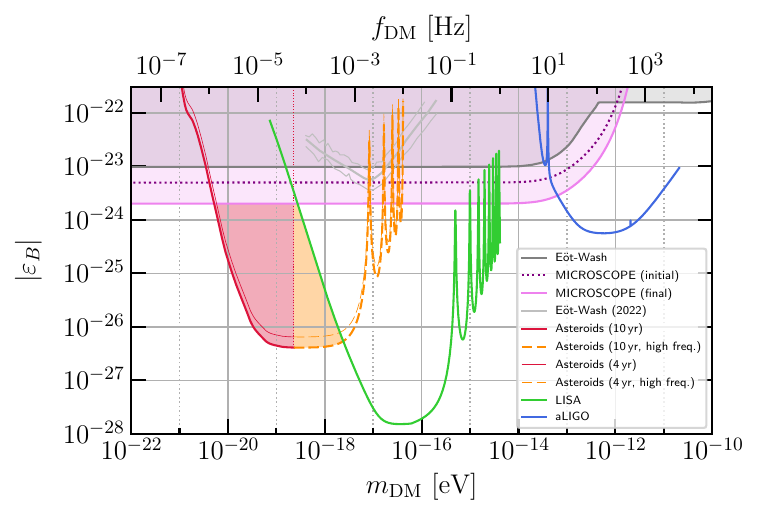}
    \includegraphics[width=0.495\textwidth]{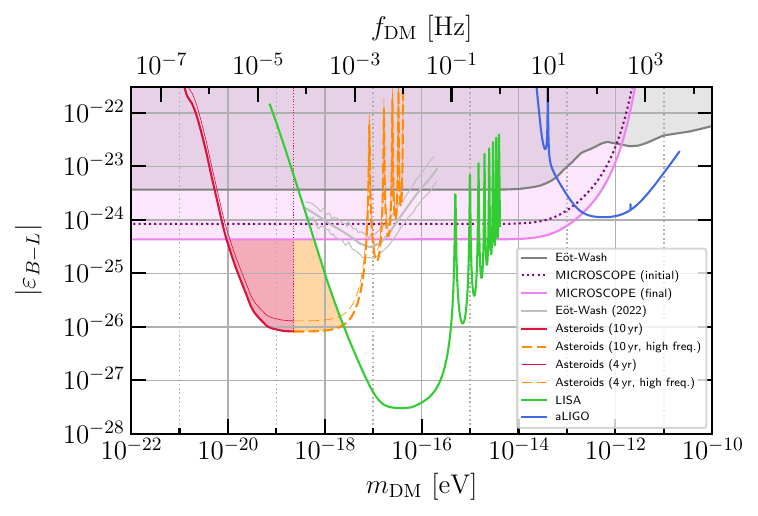}
    \caption{\label{fig:results}%
    Projected $|\varepsilon_{B}|$ [left panel] and $|\varepsilon_{B-L}|$ [right panel] sensitivity of various GW facilities/missions as labeled in the legends, assuming that the relevant dark photon is all of the local DM, $\rho_{\dm} = 0.3\,\text{GeV/cm}^3$.
    Mission parameters and sensitivities are given in \tabref{tab:parameters}.
    Our main result is the sensitivity of the asteroid-ranging proposal~\cite{Fedderke:2021kuy}, shown (for SNR of 1) as a combination of solid red and dashed orange lines. 
    The solid red part of the curves is for $f \lesssim 1/(5\,\text{hr})$, where asteroid rotational motion should not be inhibiting; the dashed orange part of the curves is for frequencies where asteroid rotational motion likely severely inhibits that proposal (see discussion in \citeR{Fedderke:2021kuy}).
    The red shaded region of parameter space could thus be accessed with the asteroid-ranging proposal (at SNR of 1), while the orange shaded region could only be accessed if noise issues associated with asteroid rotational motion could be overcome.
    We also show by the cognate thinner lines the sensitivity projections for a shorter, 4-year mission duration.
    The sensitivities we show for the LISA mission and the aLIGO facility (for comparative purposes and both also at SNR of 1) are in reasonable agreement with previous analyses~\cite{Graham:2015ifn,Pierce:2018xmy,Guo:2019ker,Morisaki:2020gui,LIGOScientificCollaborationVirgoCollaboration:2021eyz}, up to differences in assumed sensitivities and other mission/facility parameters.
    Also shown by solid medium-gray and violet lines and shaded regions, respectively, are the 95\%-confidence static EP-test constraints (which do not depend on assuming the dark photon is all, or any, of the local dark matter) from the E\"ot-Wash experiments~\cite{Schlamminger:2007ht,ADELBERGER2009102,Wagner:2012ui} and from the final results of the MICROSCOPE mission~\cite{PhysRevLett.129.121102} (re-cast using the method detailed in \citeR{PhysRevLett.120.141101}; see discussion in text).
    The dotted purple line is the cognate limit obtained from the initial MICROSCOPE results~\cite{Touboul:2017grn}; see also \citeR[s]{Fayet:2017pdp,Morisaki:2020gui}.
    Recent E\"ot-Wash results [labeled `E\"ot-Wash (2022)'] from \citeR{Shaw:2021gnp} (`field along Z' limits) are shown as light-gray lines, with the thick line representing the approximate exclusion limit averaged over nearby masses, and the limit at any specific mass fluctuating sharply within approximately the envelope denoted by the thin lines.
    }
\end{figure*}
%%%%%%%%%%%%%%%%%%%%%%%%%%%%%%%%%%%

In \figref{fig:results}, we show $\varepsilon_S$ sensitivity results (at SNR of 1) separately for $S\in\{B,B-L\}$, taking $Q_{B}/M \sim 1/\mu_a$ and $Q_{B-L}/M = (1-Z/A)/\mu_a\sim 1/(2\mu_a)$ with $\mu_a$ the atomic mass unit, and assuming the parameters and sensitivities exhibited in \tabref{tab:parameters}.
These results assume in each case that the relevant dark photon constitutes all of the local DM and we take $\rho_{\dm} = 0.3\,\text{GeV/cm}^3$ and $v_{\dm} \sim 220\,\text{km/s}$~\cite{Evans:2018bqy}; sensitivity would degrade as $\varepsilon_S \propto \sqrt{ \rho_{\dm} / \rho_V }$ for sub-component DM with energy density $\rho_V \leq \rho_{\dm}$.

Our main results are the $\varepsilon_S$ sensitivity curves for the asteroid-ranging proposal of \citeR{Fedderke:2021kuy}.
We show projected sensitivities for $f\lesssim 1/(5\,\text{hr})$ [$m_{\dm} \lesssim 2\times 10^{-19}\,$eV] in solid red, and results at higher frequencies in dashed orange; the high-frequency projections are in the region where asteroid rotational motion is expected to severely inhibit that mission proposal~\cite{Fedderke:2021kuy}. 
Note that, as discussed in detail in \appref{app:AveragingComment}, and in light of the fact that DPDM coherence times exceed the assumed mission duration for the interesting DPDM mass range for this detector, we have conservatively degraded the asteroid-detector sensitivity curves by a factor of 2 relative to those that would be obtained with the fully angularly averaged results at \eqrefRange{eq:hrms}{eq:HcalDefn}.
This accounts for an incomplete effective angular averaging over DPDM field orientations by the time-evolving baseline orientations given typical asteroid orbits, which would allow an unfortunate random but fixed DPDM field orientation out of the plane of the ecliptic to reduce the signal size.

The projected DPDM sensitivity for this mission indicates a significant reach into new parameter space at frequencies below the LISA~\cite{Babak:2021mhe} band for the $B-L$ case for DPDM in the mass range $9 \times 10^{-21}\,\text{eV} \lesssim m_{\dm} \lesssim 2 \times 10^{-19}\,\text{eV}$ and for the $B$ case for the DPDM mass range $5 \times 10^{-21}\,\text{eV} \lesssim m_{\dm} \lesssim 2 \times 10^{-19}\,\text{eV}$. 
Peak sensitivities, at $m_{\dm} \sim 2\times 10^{-19}\,$eV, exceed final MICROSCOPE bounds~\cite{PhysRevLett.129.121102} by a factor of $\sim 50$ for $\varepsilon_{B-L}$, and by a factor of $\sim 500$ for $\varepsilon_{B}$.
Sensitivity to additional new parameter space not to be otherwise probed by LISA would be possible at higher masses, up to $m_{\dm} \sim 2\times 10^{-18}\,$eV, only if noise issues arising from asteroid rotational motion could be overcome.

%%%%%%%%%%
\begin{table}[t]
    \begin{ruledtabular}
    \begin{tabular}{lllp{3cm}}
         Detector   & $L$ [m] & $T$ [yr] & Sensitivity from\\ \hline
         aLIGO      &   $\phantom{0.}4\times 10^3$      &   4   &   \citeR{LIGOcurve}    \\
         LISA       &   $2.5\times 10^9$    &   4   &   \citeR{Babak:2021mhe}; Eq.~(114)  \\
         Asteroids  &   $1.5\times 10^{11}$ ($\approx 1\,\text{AU}$)    & 10     &   \citeR{Fedderke:2021kuy}; `smoothed envelope 2' curve from Fig.~9
    \end{tabular}
    \end{ruledtabular}
    \caption{\label{tab:parameters}%
        Parameters assumed for the sensitivity curves shown in \figref{fig:results}. 
        $L$ is the baseline length and $T$ is the assumed total mission or observation duration. 
        The `sensitivity from' column gives the reference and location from which the relevant strain-sensitivity curve (or cognate) is taken.}
\end{table}
%%%%%%%%%%%%

Our results for LISA and aLIGO~\cite{LIGOcurve}, shown for comparative purposes, are in reasonable agreement with results of previous works that have considered one or both of those missions/facilities for this purpose~\cite{Graham:2015ifn,Pierce:2018xmy,Guo:2019ker,Morisaki:2020gui,LIGOScientificCollaborationVirgoCollaboration:2021eyz}; see \citeR{Morisaki:2020gui} for sensitivity projections for other future GW detectors operating between the LISA and aLIGO bands, and in the aLIGO band.

We also show in \figref{fig:results} existing 95\%-confidence constraints from static EP tests.
These bounds rely only on the existence of a massive $U(1)$ gauge boson weakly coupled to the relevant SM current; they do not need to assume that this dark photon is all (or indeed, any fraction) of the local DM. 
Constraints from the E\"ot-Wash torsion-balance experiments~\cite{Schlamminger:2007ht} on $\epsilon_B$ are taken from graphical results in \citeR{ADELBERGER2009102}, and on $\epsilon_{B-L}$ from graphical results in \citeR{Wagner:2012ui} (which are slightly stronger and updated as compared to those in \citeR{ADELBERGER2009102}).
Also shown are recent E\"ot-Wash results from \citeR{Shaw:2021gnp} (we show their `field along Z' limits).
The constraints from the MICROSCOPE mission in \figref{fig:results} are obtained as follows: we utilize the constraints on the Ti-Pt E\"otv\"os parameter $\eta(\text{Ti},\text{Pt})$ that are given in either \citeR{Touboul:2017grn} (`initial') or \citeR{PhysRevLett.129.121102} (`final'), and follow the procedures in \citeR{PhysRevLett.120.141101} to recast these limits as constraints on $\varepsilon_{S}$.
We opt to add statistical and systematic errors on $\eta \equiv \eta(\text{Ti},\text{Pt})$ in quadrature, and we report exclusions at $2\sigma$ significance, accounting for the (signed) non-zero reported central values $\hat{\eta}$ given in \citeR[s]{Touboul:2017grn,PhysRevLett.129.121102}; that is, we report a value of $\varepsilon_{S}$ to be excluded if it results%
%%%%%%%%%%%%
\footnote{\label{ftnt:signs}%
    Recall that $U(1)$ exchange creates a repulsive potential for like-sign charges, leading to $\alpha<0$ in Eqs.~(5) and~(6) of \citeR{PhysRevLett.120.141101}.} %
%%%%%%%%%%%%
in $| \eta(\varepsilon) - \hat{\eta} | > 2\sigma_\eta$.
To cross-check our recasting of the `initial' results, we also separately looked at the corresponding constraints on the parameter $|\alpha_{S}| \equiv \varepsilon_{S}^2\alpha_{\tsc{em}} / ( G_N \mu_a^2 )$ defined in \citeR{PhysRevLett.120.141101}.%
%%%%%%%%%%%%
\footnote{\label{ftnt:offsetIrrelevant}%
   For the initial MICROSCOPE results, $\hat{\eta} / (\sigma_\eta) \ll 1$, so limits are nearly symmetric about $\alpha_{S} = 0$; this is not true for the  final results, and we use the correctly signed value of $\alpha_{S}$ (here, negative) to extract limits on $\varepsilon_{S}$.} %
%%%%%%%%%%%%
We find here results that disagree mildly by overall factors with the graphical results shown in \citeR{PhysRevLett.120.141101} itself: our limits are a factor of $\sim 3$ weaker for $|\alpha_B|$ and a factor of $\sim 2$ weaker for $|\alpha_{B-L}|$; in both cases, however, the functional form of the limits as $\lambda\equiv 1/m_{\dm}$ changes is however reproduced exactly.
The origin of these overall-factor discrepancies is not readily apparent to us (a small difference%
%%%%%%%%%%%%
\footnote{\label{ftnt:uncerts}%
    \citeR{Touboul:2017grn} gives $\sigma^\eta_{\text{stat}} \approx \sigma^\eta_{\text{syst}} \approx 9\times 10^{-15}$ with the statistical error given at $1\sigma$.
    \citeR{PhysRevLett.120.141101} quotes a combined 2$\sigma$ uncertainty of $2\sigma^\eta_{\text{tot}} \approx 27 \times 10^{-15}$.
    We use a combined $2\sigma$ uncertainty of $2\sigma^\eta_{\text{tot}} \approx 25 \times 10^{-15}$, consistent with \citeR{Fayet:2017pdp}.} %
%%%%%%%%%%%%
in the treatment of the addition of statistical and systematic uncertainties cannot account for it).
Moreover, our recasting agrees well with an independent recasting that was performed by \citeR{Fayet:2017pdp} and also reported in \citeR{Morisaki:2020gui}.
We merely make the reader aware of this issue and point out that the limits on $\varepsilon_{S}$ that we report for the initial MICROSCOPE results are more conservative by factors of~$\sim 1$--$2$ than those that can be inferred from the graphical results of \citeR{PhysRevLett.120.141101}.
Per our recasting, the recently released final MICROSCOPE results~\cite{PhysRevLett.129.121102} imply limits on $\varepsilon_S$ a factor of $\sim 2$ stronger than the initial results~\cite{Touboul:2017grn}.\\

In this paper we have the evaluated the sensitivity of the asteroid-ranging GW-detection proposal recently advanced in \citeR{Fedderke:2021kuy} to dark-photon dark matter that is coupled to the $B$ or $B-L$ charges of the SM.
Our results show that this GW detector could also probe significant new regions of parameter space for these DM candidates.
This extends the science case for the development of this class of $\mu$Hz GW detectors.

%%%%%%%%%%%%%%%%%%%%%%%%%%%%%%%%%%%%%%%%%%%%%%%%%%%%%%%%%%%%%%%%%%%%%%%%%%%%%%%%%%%%%%%%%%
\acknowledgments
%%%%%%%%%%%%%%%%%%%%%%%%%%%%%%%%%%%%%%%%%%%%%%%%%%%%%%%%%%%%%%%%%%%%%%%%%%%%%%%%%%%%%%%%%%
We thank Peter W.~Graham, David E.~Kaplan, and Surjeet Rajendran for useful discussions.  
 
This work was supported by the U.S.~Department of Energy (DOE), Office of Science, National Quantum Information Science Research Centers, Superconducting Quantum Materials and Systems Center (SQMS) under Contract No.~DE-AC02-07CH11359. 
This work was also supported by the Simons Investigator Award No.~827042~(P.I.: Surjeet Rajendran).

%%%%%%%%%%%%%%%%%%%%%%%%%%%%%%%%%%%%%%%%%%%%%%%%%%%%%%%%%%%%%%%%%%%%%%%%%%%%%%%%%%%%%%%%%%
\appendix
\section{Detector correction factors \texorpdfstring{$k_{\text{det}}(f)$}{}}
\label{app:kdetFactors}
%%%%%%%%%%%%%%%%%%%%%%%%%%%%%%%%%%%%%%%%%%%%%%%%%%%%%%%%%%%%%%%%%%%%%%%%%%%%%%%%%%%%%%%%%%
In this appendix, we specify the detector-specific correction factors $k_{\text{det}}(f)$ that we apply in \eqref{eq:sensitivity}.

For the asteroid-ranging proposal of \citeR{Fedderke:2021kuy}, the detector strain response $h_{\tsc{gw}}(t)$ to a plane GW with frequency $f_{\tsc{gw}} = \omega_{\tsc{gw}}/(2\pi)$ and amplitude $h_0$ was in principle assumed to be that of an optimally oriented $L \sim 1\,$AU single baseline:
%%%%%%
\begin{align}
h_{\tsc{gw}}(t) = \frac{h_0}{2} \sinc(\omega_{\tsc{gw}} L) \cos\lb[ \omega_{\tsc{gw}} t + \beta \rb]\:,
\label{eq:AstStrainResp}
\end{align}
%%%%%%
where $\beta$ is a phase.
No averaging over baseline orientation, GW propagation direction, or GW polarization (hereinafter, `geometrical averaging') was accounted for in \citeR{Fedderke:2021kuy}.
The time average of the square of the optimally oriented $h_{\tsc{gw}}(t)$ is $\langle h_{\tsc{gw}}^2 \rangle = (1/8) h_0^2 \,\sinc^2(\omega_{\tsc{gw}} L)$, and the characteristic-strain sensitivity $h_c(f)$ in \citeR{Fedderke:2021kuy} was expressed%
%%%%%%%%%%%%
\footnote{\label{ftnt:hcitoh0}%
    In \citeR{Fedderke:2021kuy}, see, e.g., Eq.~(5), footnote 9, and Eqs.~(87) and (88) as well as the discussion that follows the latter.
    } %
%%%%%%%%%%%%
such that a GW of amplitude $h_0 \sim h_c / \sqrt{f_{\tsc{gw}}T}$ is detectable at SNR of 1 given an observation time $T$.
Therefore, the detectable amplitude of the mean-square detector-strain response (at SNR of 1) is
%%%%%%
\begin{align}
\langle h^2 \rangle = \frac{ S_n(f) }{ 2 T } \times \frac{1}{4} \sinc^2(\omega_{\tsc{gw}} L)\:,
\end{align}
%%%%%%
where we used that $f S_n(f) \equiv [h_c(f)]^2$.
By comparison with \eqref{eq:sensitivity} [and noting that we have assumed $T_{\text{coh}}\rightarrow\infty$ in this discussion], we see that $k^{\text{ast}}_{\text{det}} = \sinc^2(\omega_{\tsc{gw}} L)/4$.
However, the $\sinc$ factor was treated only approximately in \citeR{Fedderke:2021kuy}, with the $h_c$ curve simply degraded by a factor $f/f_\star$ on $f \gtrsim f_{\star} \equiv 1/(\pi L)$; for consistency, we must adopt the same approximation here. 
The correction factor to be applied to the results of \citeR{Fedderke:2021kuy} is thus
%%%%%%
\begin{align}
    k_{\text{det}}^{\text{ast}} = \frac{1}{4} 
        \times 
            \begin{cases}
                1 & f \leq 1/(\pi L) \\
                (\pi f L)^{-2} & f > 1/(\pi L)
            \end{cases}\:.
\end{align}
%%%%%%
Note that a high-frequency signal-response suppression factor cognate to the $\sinc$ factor in \eqref{eq:AstStrainResp} operates to parametrically suppress the $\langle h_1^2 \rangle$ signal for the DPDM case [\eqref{eq:h1rms}], but does not operate to parametrically suppress $\langle h_2^2 \rangle$ [\eqref{eq:h2rms}]; see also \citeR{Morisaki:2020gui}.

It was pointed out in \citeR{Morisaki:2020gui}, and we agree, that were a geometrical-averaging factor included in a GW detector sensitivity curve, a correction must be applied to remove that geometrical factor before employing that sensitivity curve in \eqref{eq:sensitivity}; otherwise, a double counting of geometrical factors would be involved when comparing to the DPDM signal at \eqrefRange{eq:hrms}{eq:HcalDefn}, which itself has been geometrically averaged over DPDM polarization states and propagation directions.
As noted above, no geometrical averaging was performed for the asteroids detector, so no such correction is needed in that case.

Let us now discuss the aLIGO case.
Conventionally, aLIGO reports strain-sensitivity curves with geometrical-averaging factors already removed (they are assigned instead to the source); see, e.g., discussions in \citeR[s]{Babak:2021mhe,Robson:2018ifk}.
As such, no correction for geometrical averaging is required for aLIGO curves.
Moreover, the cognate of the explicit factor of $1/2$ in \eqref{eq:AstStrainResp} is absent for aLIGO since, at optimum orientation, the strains in the two perpendicular baselines are of opposite sign and equal magnitude, and thus add in the difference signal.
In principle, we should still make a correction for a high-frequency suppression factor $\sim \sinc^{2}(\omega_{\textsc{gw}}L)$ [note: the transfer functions for the aLIGO FP cavities are present in both GW and DPDM cases, and require no additional correction~\cite{Morisaki:2020gui}; see footnote \ref{ftnt:FPcavity}].
However, aLIGO only accesses frequencies $f \lesssim 5\,\text{kHz}$, where $\sinc^2(\omega_{\textsc{gw}}L)\gtrsim 0.94$; already at $f\sim \text{kHz}$, the correction is sub-percent.
To the level of accuracy we work, and for the frequencies we consider, we therefore ignore any correction for this effect.
As such, the aLIGO correction factor is taken to be
%%%%%%
\begin{align}
    k_{\text{det}}^{\text{aLIGO}} = 1\:.
\end{align}
%%%%%%

On the other hand, LISA sensitivity curves are conventionally degraded by a geometrical-averaging factor~\cite{Babak:2021mhe,Robson:2018ifk}. 
Removing this degradation as required~\cite{Morisaki:2020gui}, while also accounting for an $\order{1}$ numerical factor related to LISA arm configurations, inserts a factor of $3/20$ in $k_{\text{det}}^{\text{LISA}}$ (see, e.g., the discussion below Eq.~(84) in \citeR{Babak:2021mhe}).
Finally, we use the LISA `SciRD' sensitivity defined at Eq.~(114) in \citeR{Babak:2021mhe} (see \tabref{tab:parameters}), in which the high-frequency detector-response suppression is captured entirely by the factor defined there as $R(f) \equiv 1 + [f/(25\,\text{mHz})]^2$; to correct for this, we include a factor of $\lb[ R(f) \rb]^{-1}$ in $k_{\text{det}}^{\text{LISA}}$.
Overall, the LISA correction factor is taken to be
%%%%%%
\begin{align}
    k_{\text{det}}^{\text{LISA}} = \frac{3}{20 \lb( 1 + [f/(25\,\text{mHz})]^2 \rb)}\:.
\end{align}
%%%%%%.

%%%%%%%%%%%%%%%%%%%%%%%%%%%%%%%%%%%%%%%%%%%%%%%%%%%%%%%%%
\section{Comments on DPDM polarization and momentum-direction averaging}
\label{app:AveragingComment}
%%%%%%%%%%%%%%%%%%%%%%%%%%%%%%%%%%%%%%%%%%%%%%%%%%%%%%%%%

Some comments are in order on our averaging over DPDM polarization and momentum directions at \eqrefRange{eq:hrms}{eq:HcalDefn}.

One can consider two possible extreme cases for the DPDM polarization state, depending on the DPDM formation mechanism and subsequent structure-formation processing~\cite{Arias:2012az,Caputo:2021eaa}: either the DPDM polarization direction randomizes over a coherence time/length, or the DPDM polarization state is instead fixed on patches much larger than a coherence length (in the extreme limit, the DPDM polarization state could be fixed in our entire Hubble patch~\cite{Arias:2012az,Caputo:2021eaa}).
In either case, the DPDM momentum direction (i.e., local spatial gradient) will still randomize on the coherence time.

In practice, the coherence time for $m_{\textsc{dm}} \lesssim (\text{few})\times 10^{-17}\,$eV is longer than the assumed $\sim 4$--$10$\,yr mission durations we consider in this work (see \sectref{sect:results} and \tabref{tab:parameters}) so both of these two fundamental alternatives in the DPDM polarization treatment would in practice result in the DPDM polarization vector being fixed on the duration of observation we consider for such masses; for higher masses, these alternatives are of course different in practice.

It is clear that if the DPDM polarization state does randomize, and the coherence time is shorter than the mission or observation duration, then polarization averaging at \eqrefRange{eq:hrms}{eq:HcalDefn} is motivated to derive a single mission-averaged figure-of-merit sensitivity.
Likewise, if the coherence time is shorter than the mission or observation duration, an averaging over momentum directions is justified.

In the case that the polarization state is however fixed or varies only very slowly compared to assumed mission durations (long coherence time), the argument in favor of DPDM polarization and/or momentum-direction averaging would need to rely on the fact that real GW detector baselines vary over time in their orientations with respect to the DPDM polarization state and momentum direction (which are both fixed in inertial space) by large enough amounts that an effective averaging over a large range of orientation angles occurs naturally over the course of a mission/experiment.

For aLIGO, the orientation of the detector arms at the two sites in Hanford and Livingston relative to inertial space, along with the rotation of the Earth over time, prohibits the possibility that the DPDM polarization and momentum vectors can be so poorly oriented with respect to all of the baselines at all times that a significant signal suppression below the polarization- and momentum-direction-averaged result could occur.
The fully angularly averaged results at \eqrefRange{eq:hrms}{eq:HcalDefn} thus give, up to possible $\mathcal{O}(1)$ factors, a good experiment-averaged sensitivity projection, regardless of the DPDM polarization treatment or the coherence time.

The proposed Earth-trailing heliocentric LISA satellite orbits will result in a rotating triangular configuration for the constellation, lying in a plane offset by $60^\circ$ relative to the plane of the ecliptic~\cite{LISA_L3}. 
This will result in an $\mathcal{O}(1)$ fraction of relative orientations of the baselines with respect to any possible fixed DPDM polarization state and/or momentum direction being explored.
The angularly averaged results at \eqrefRange{eq:hrms}{eq:HcalDefn} thus again give, up to possible $\mathcal{O}(1)$ factors, a good mission-averaged sensitivity projection, regardless of the DPDM polarization treatment, or coherence time.

The case of the asteroid-based detector is the more nuanced and requires a correction.
The types of asteroids that one could utilize as test masses in the proposal of \citeR{Fedderke:2021kuy} typically have mildly elliptical orbits with different semi-major axes, and are typically inclined to the plane of the ecliptic by $\mathcal{O}(10^{\circ})$, although some are more inclined, and some less; see Table I of \citeR{Fedderke:2021kuy}.
(Note that we ignore baseline length variations completely in this work.)
For a typical choice of asteroids then, the baselines defined by their motion will naturally explore all orientations in the plane of the ecliptic, but only a limited set of orientations out of the plane of the ecliptic, varying perhaps by $\pm 10^{\circ}$ over orbital/mission timescales.
For interesting DPDM mass ranges, coherence times are such that the asteroid detector is always to be considered in the regime where the DPDM polarization state and momentum direction are fixed to some random direction over the whole mission duration we consider, regardless of the fundamental underlying DPDM polarization treatment.  
As such, it is in principle possible that the DPDM polarization state and/or momentum direction can be unluckily oriented mostly perpendicular to the plane of the ecliptic, resulting in a suppression of the signal size below the angular average.

Recalling from the discussion in the paragraph between \eqref[s]{eq:HcalDefn} and (\ref{eq:cGeomA}) that the $h_1$ signal is dominant in the interesting mass range for the asteroid-based detector, the  $\varepsilon$ sensitivity for the un-averaged case is degraded (or enhanced) by a factor of $\zeta \equiv (3\cos^2\theta)^{-1/2}$ compared to the averaged case, where $\theta$ is the angle between the baseline and DPDM polarization state.
For favorably aligned orientations, $1/\sqrt{3}\leq \zeta \leq 1$, and the unaveraged sensitivity is enhanced; for near-orthogonal orientations, the unaveraged sensitivity is degraded ($\zeta \geq 1$).

In the worst case for the asteroid-based detector, given that typical baselines would explore varying orientations within an angle on the order of $\pm 10^{\circ}$ from the plane of the ecliptic throughout the two asteroid orbits, one can estimate the signal degradation by taking $\theta \sim 80^{\circ}$, resulting in a degradation by a factor of $\zeta \sim 3$; for a $\pm 5^\circ$ offset, a factor of $\zeta \sim 7$ degradation is to be expected.

However, these are overly pessimistic assumptions: following discussions in~\citeR[s]{Arias:2012az,Caputo:2021eaa}, a better assumption is to set the orientation angle to the worst case one would expect at the same level of confidence one is making projections.
Because our sensitivity projections are all at SNR of 1, corresponding to 68\% confidence, the corresponding worst-case value is $\cos^2\theta \sim (0.32)^2$, assuming a flat prior on $\cos\theta$, in line with the Jacobian for spherical co-ordinates.
This would lead to a degradation in sensitivity by only a factor of $\zeta \sim 2$ as compared the sensitivity projections based on the fully-averaged result.
As a result, we have degraded the sensitivity curves shown in \figref{fig:results} for the asteroid-based detector proposal by a factor of 2 from the result that would be obtained from the fully angularly averaged expressions at \eqrefRange{eq:hrms}{eq:HcalDefn}, as we mention in \sectref{sect:results}.
Note that we apply no correction for LISA or aLIGO sensitivity curves.

While it is plausible that one could possibly select one of the asteroids for the mission proposed in \citeR{Fedderke:2021kuy} with a much larger inclination angle, perhaps up to $\sim 40^\circ$ (in which case only a small, $\mathcal{O}(1)$ degradation would be expected in DPDM sensitivity), this may not be possible for other reasons having to do with asteroid selection criteria, so it is not guaranteed one could avoid the factor-of-2 suppression in this way.

We also note finally that modulations of the signal will occur in all of the above cases by virtue of relative orientation considerations if the DPDM polarization state is fixed on long timescales, and those would be important to consider for an actual signal search in data. 
However, the angularly averaged result (appropriately corrected, where necessary, as discussed in this appendix) gives an appropriate figure of merit for sensitivity projections.

%%%%%%%%%%%%%%%%%%%%%%%%%%%%%%%%%%%%%%%%%%%%%%%%%%%%%%%%%%%%%%%%%%%%%%%%%%%%%%%%%%%%%%%%%%
%%%%%%%%%%%%%%%%%%%%%%%%%%%%%%%%%%%%%%%%%%%%%%%%%%%%%%%%%%%%%%%%%%%%%%%%%%%%%%%%%%%%%%%%%%
\bibliographystyle{JHEP}
\bibliography{references.bib}
%%%%%%%%%%%%%%%%%%%%%%%%%%%%%%%%%%%%%%%%%%%%%%%%%%%%%%%%%%%%%%%%%%%%%%%%%%%%%%%%%%%%%%%%%%
%%%%%%%%%%%%%%%%%%%%%%%%%%%%%%%%%%%%%%%%%%%%%%%%%%%%%%%%%%%%%%%%%%%%%%%%%%%%%%%%%%%%%%%%%%

%%%%%%%%%%%%%%%%%%%%%%%%%%%%%%%%%%%%%%%%%%%%%%%%%%%%%%%%%%%%%%%%%%%%%%%%%%%%%%%%%%%%%%%%%%
%%%%%%%%%%%%%%%%%%%%%%%%%%%%%%%%%%%%%%%%%%%%%%%%%%%%%%%%%%%%%%%%%%%%%%%%%%%%%%%%%%%%%%%%%%
%%%%%%%%%%%%%%%%%%%%%%%%%%%%%%%%%%%%%%%%%%%%%%%%%%%%%%%%%%%%%%%%%%%%%%%%%%%%%%%%%%%%%%%%%%
%%%%%%%%%%%%%%%%%%%%%%%%%%%%%%%%%%%%%%%%%%%%%%%%%%%%%%%%%%%%%%%%%%%%%%%%%%%%%%%%%%%%%%%%%%
\end{document}